\documentclass[12pt,preprint,psfig]{aastex}
\usepackage{emulateapj5}

\def\ltsima{$\; \buildrel < \over \sim \;$}

\def\kms     {km~s$^{-1}$}
\def\Msun    {M$_\odot$}
\def\lsim{\lower.5ex\hbox{\ltsima}}
\def\gtsima{$\; \buildrel > \over \sim \;$}
\def\gsim{\lower.5ex\hbox{\gtsima}}
\begin{document}

\title{Constraints on the Proper Motion of the Andromeda Galaxy Based on
the Survival of Its Satellite M33}

\author{Abraham Loeb, Mark J. Reid}

\affil{Harvard-Smithsonian Center for Astrophysics, 60 Garden Street, Cambridge, MA 02138, USA}

\author{Andreas Brunthaler}

\affil{Joint Institute for VLBI in Europe, Postbus 2, 7990 AA
  Dwingeloo, The Netherlands}

\author{Heino Falcke}

\affil{ASTRON, Postbus 2, 7990 AA Dwingeloo, The Netherlands\\
 Department of Astrophysics, Radboud Universiteit
              Nijmegen, Postbus 9010, 6500 GL Nijmegen,\\ The Netherlands} 

\begin{abstract}

A major uncertainty in the dynamical history of the local group of galaxies
originates from the unknown transverse speed of the Andromeda galaxy (M31)
relative to the Milky Way.  We show that the recent VLBA measurement of the
proper motion of Andromeda's satellite, M33, severely constrains the
possible values of M31's proper motion.  The condition that M33's stellar
disk will not be tidally disrupted by either M31 or the Milky Way over the
past 10 billion years, favors a proper motion amplitude of $100\pm20 ~{\rm
km~s^{-1}}$ for M31 with the quadrant of a negative velocity component
along Right Ascension and a positive component along Declination strongly
ruled-out. This inference can be tested by future astrometric measurements
with SIM, GAIA, or the SKA.  Our results imply that the dark halos of
Andromeda and the Milky Way will pass through each other within the next
5--10 billion years.
\end{abstract}

%\keywords{Local Group -- galaxies: individual (M31, M33) -- galaxies:
%interactions -- cosmology: dark matter -- astrometry}

\section{Introduction}

The local group of galaxies provides the nearest laboratory for the
dynamics and evolution of galaxies. Its two dominant galaxies, the Milky
Way and Andromeda (M31) are separated by a distance of $\sim 700$ kpc and
are moving towards each other with a radial velocity of $-117~{\rm
km~s^{-1}}$ \citep{BT}.  The unknown transverse velocity of M31 relative to
the Milky Way introduces an ambiguity into the dynamical history of the
local group and the likely dark-matter mass of its components
\citep{Peebles1,Peebles2}. The system potentially provides a local example
for the dynamics that leads more generally to the formation of elliptical
galaxies out of the merger of two disks \citep{Barnes}.

The estimated ages of stars within the disks of the Milky Way and Andromeda
imply that these disks formed $\sim 10$ billion years ago, and their small
scale height suggests that they had not experienced any substantial merger
activity since then \citep{Wyse1,Wyse2,Pel,Toth}.

Recent VLBA observations of two H$_2$O masers on opposite sides of
Andromeda's satellite, M33, led to a geometric determination of its 
distance $730 \pm 168$ kpc and total velocity of $190\pm 59~{\rm
km~s^{-1}}$ relative to the Milky Way \citep{Bru}. Unfortunately, no such
masers were found in Andromeda, despite extensive search efforts (e.g. Imai
et al. 2001).  M33 has a thin \ion{H}{1}/stellar disk that extends out to a
radius of $\sim 10$ kpc \citep{Corb}.  Here we use the existence of this
disk and its measured position and bulk velocity in three-dimensions, to
constrain the transverse velocity of Andromeda.  In particular, we use the
constraints that: {\it (i)} M33's stellar disk was not disrupted by tidal
interactions with the Milky Way and Andromeda over the past 10 billion
years; and {\it (ii)} these galaxies were closer to each other more than 10
billion years ago as participants in the expansion of the universe.  In \S
2 we describe our prescription for the mass profiles of the galaxy halos
under consideration and in \S 3 we present our numerical results. Finally,
\S 4 summarizes our main conclusions.

\section{Halo Mass Profiles}

We would like to constrain the velocity vector of M31 based on the
condition that the stellar disk of M33 (as represented by test particles on
circular orbits at radii $\la 10$ kpc from the center of M33) was not
disrupted during the past $\sim 10$ billion years.  
%Together with the measured radial velocity of M31 relative to the Milky
%Way, this would set a lower limit on the mass of the Milky Way galaxy in
%order for M31 to be gravitationally bound to the local group.
For simplicity, we assume a spherical distribution of enclosed mass,
$M(r)$, within each of the galaxy halos under consideration (M31, M33, and
the Milky Way) and follow orbits of test particles in them. In order to
solve for the orbits of these test particles we only need to specify the
gravitational acceleration as a function of radial distance $r$ from the
center of these halos, ${\vec g}=g_r(r) {\hat e}_r$, where
\begin{equation}
g_r= -{GM(r)\over r^2}= -{v^2(r)\over r} ,
\end{equation}
with $v(r)$ being the circular velocity as function of radius within these
halos. 

Our test-particle approach ignores the effect of dynamical friction on the
orbit of M33 inside the halo of M31. The additional effect of dynamical
friction can only strengthen our constraints since it would bring M33
closer to the core of M31, where it would be more vulnerable to tidal
disruption. Hence, our simple test-particle approach provides 
conservative constraints on the allowed range of possible orbits for the
M33/M31 system. For an elaborate discussion on the impact of dynamical
friction on the survival of satellite halos, see Taffoni et al. (2003) and
references therein.

The simplest model of galaxy halos assumes a flat rotation curve,
$v=const$, out to a truncation radius $r_t$. Since all three galaxies under
consideration have \ion{H}{1} disks, the values of $v$ can be inferred from
their corresponding kinematic data, namely $\sim 110$, 220, and 250~${\rm
km~s^{-1}}$ for M33, the Milky Way, and M31, respectively.  The truncation
radius is related to the total mass of the halo, $M_{\rm t}$, through the
expression $r_t= {GM_{\rm t}/v^2}$.  For M33, $r_t$ is dictated by the
tidal effect of M31. Given an orbit of M33, we find $r_t$ for M33 by
searching for the largest radius at which a test particle initially on a
circular orbit around M33 would remain bound to it.

For M31 and the Milky Way, it is natural to associate $r_t$ with the virial
radius, $r_v$, inside of which the mean halo density is $\sim 200$ times
larger than the mean density of the universe 
%(according to the spherical collapse model) 
when these galaxies formed.  More precisely, in a flat universe with
density parameters $\Omega_m$ in matter and $\Omega_\Lambda = 1-\Omega_m$
in a cosmological constant, the density of the halo in units of the
critical density at the collapse redshift $z$ of the halo is (see \S 2.3 in
Barkana \& Loeb 2001 for more details),
\begin{equation}
\Delta_c=18\pi^2+82d-39 d^2,
\end{equation}
where $d=\Omega_m^z-1$ and
\begin{equation}
\Omega_m^z={\Omega_m(1+z)^3\over \Omega_m(1+z)^3+\Omega_\Lambda}.
\end{equation}
A halo of virial mass $M_v$ that formed at a redshift $z$ has a physical
virial radius of
\begin{equation}
r_v=84.5 \times M_{12}^{1/3}
[f(z)]^{-1/3}
\left({1+z\over 2}\right)^{-1} h^{-1}~{\rm kpc},
\label{eq:r_v}
\end{equation}
and a corresponding circular velocity at the virial radius,
\begin{equation}
v_v=225.5 \times M_{12}^{1/3} [f(z)]^{1/6}\left({1+z\over
2}\right)^{1/2}~{\rm km~s^{-1}}~.
\label{eq:v_v}
\end{equation}
Here $M_{12}\equiv ({M_v/10^{12} h^{-1}M_\odot})$, $h$ is the present-day
value of the Hubble constant in units of $100~{\rm km~s^{-1}Mpc^{-1}}$, and
$f(z)\equiv [(\Omega_m/\Omega_m^z)(\Delta_c/18\pi^2)]$.  Our choice for the
virialization redshift of M31, M33, and the Milky Way is $z\sim 1$
\citep{Wyse1,Wyse2}.  Nuclear dating of the thin Galactic disk implies an
age of $\sim 8.3\pm 1.8$ Gyr corresponding to a formation redshift $z\sim
1$ (Peloso et al. 2005), as expected from other considerations for spiral
galaxies (Hammer et al. 2005). The total halo masses of the Milky Way and
M31 were also constrained directly in the literature
\citep{BT,Gott,Battaglia}.  We parametrize the masses of M31 and the
Milky Way in units of $3.4$ and $2.3 \times10^{12}$ \Msun\ respectively,
and adopt the WMAP values for the cosmological parameters, namely
$\Omega_m=0.27$, $\Omega_\Lambda=1-\Omega_m=0.73$, and $h=0.71$ (Spergel et
al. 2003).

To refine our constraints, we adopt a halo profile that follows from
cosmological simulations of galaxy formation.  Navarro, Frenk, \& White
(1997; hereafter NFW) found an analytic fit to results from their
cosmological N-body simulations of galaxy halos in a cold dark matter
cosmology.  The density profile in their model has an asymptotic radial
dependence of $r^{-3}$ at large distances and $r^{-1}$ at small
distances. The model has two free parameters: one is the transition radius
that separates these different power-law dependencies, $r_s$, and the
second is the so-called ``concentration parameter'', $c$, which denotes the
ratio between the virial radius $r_v$ and $r_s$.

The radial gravitational acceleration in the NFW model is given by
\begin{equation}
g_r= -{GM_v \over r^2}{\ln (1+cx)-(cx)/(1+cx) \over 
[\ln (1+c) -c/(1+c)]}
\end{equation}
where $x=r/r_v$ with $r_v$ given by equation (\ref{eq:r_v}).  The
present-day concentration parameter is correlated with the halo collapse
redshift (Wechsler et al. 2002).  

\section{Numerical Results}

\begin{figure*}[t]
\epsscale{1} 
\plotone{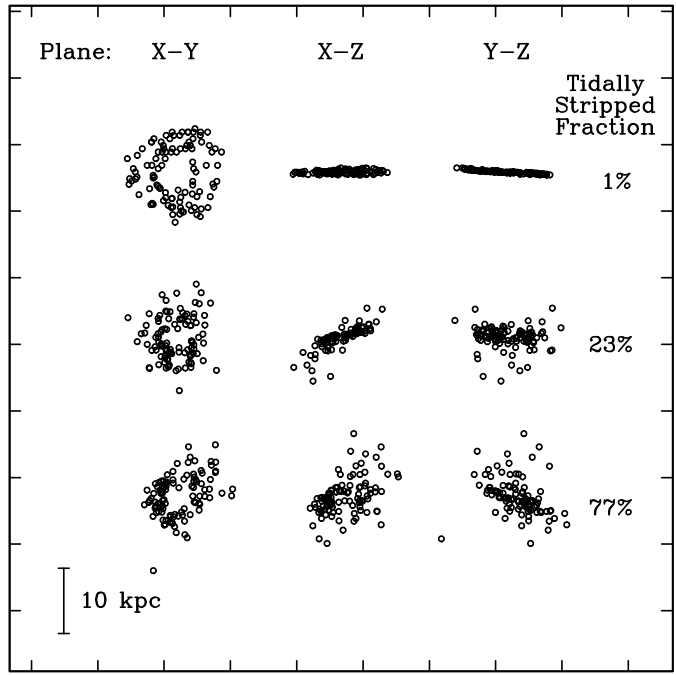}
\caption{
\label{fig1}
Three examples of tidal stripping and heating of the stellar disk of M33.
The locations of stars at the end of the N-body simulation are shown
projected onto three orthogonal planes, as labeled at the top of the
figure.  These examples all used the full dark matter halos ($m_f=1$) and
simulated histories over the past 10 Gyr.  A 10 kpc scale is shown in the
bottom left corner.  {\it Top:} Simulation with $(V_{ \alpha\cos\delta
},V_\delta ) = (100,-100)$~\kms for M31, which resulted in essentially no
tidal stripping (tidally stripped fraction of $\sim 1$\%).  {\it Middle:}
Simulation with $(V_{\alpha\cos\delta },V_\delta ) = (-50,-50)$~\kms for
M31, which resulted in moderate tidal stripping (23\%).  {\it Bottom:}
Simulation with $(V_{\alpha\cos\delta},V_\delta )= (0,0)$~\kms for M31,
which resulted in heavy tidal stripping and heating (77\%).  }
\end{figure*}

Our goal is to examine a wide range of trial proper motions for M31,
compute the history of the M33/M31/Milky Way system, and determine which
proper motions lead to the survival of M33's stellar disk. Our numerical
simulations used S. Aarseth's NBODY0
code\footnote{http://bima.astro.umd.edu/nemo/; see also
http://www.ast.cam.ac.uk/\~sverre/web/pages/home.htm}, modified to allow
the extended dark-matter potentials associated with the galaxies.  M31 and
the Milky Way were treated as extended rigid bodies while the disk and halo
of M33 were modelled as a collection of test particles moving in the
combined, time-dependent potential of M31, the Milky Way, and M33.  We used
a Galactic Cartesian coordinate system in which the Milky Way is in the
$(x,y)$ plane, with the Galactic Center at $(8,0,0)$~kpc, $y$ is toward
Galactic rotation, and $z$ is toward the North Galactic Pole.

The calculations were initialized to match today's positions and motions of
M33, M31, and the Milky Way.  While we have a measured proper motion for
M33, none exist for M31.  Therefore, we assigned trial values for the
proper motion for M31 in the eastward ($\alpha\cos{\delta}$) and northward
($\delta$) directions and then converted them to the Galactic Cartesian
frame.  The observed radial velocity of M31 relative to the center of the
Milky Way is $-117$~\kms, which converted to the Galactic Cartesian frame
yields $(56,-93,43)$~\kms.  The radial and trial proper motions were
combined to give the full velocity vector of M31 at the present epoch.  For
the distances of M31 and M33 from the Milky Way we adopt the respective
values of 786 kpc and 794 kpc \citep{Mc1,Mc2}.

M31 and the Milky Way were assigned total masses of $3.4 m_f$ and $2.3
m_f\times10^{12}$ \Msun\ and truncation radii $r_t$ of $235 m_f$ and $207
m_f$ kpc respectively (consistently with their measured rotation curves for
the isothermal sphere model).  The scaling parameter $m_f$ was used to test
the sensitivity of our results to the total dark matter masses (for recent
empirical constraints, see Battaglia et al. (2005) and references therein).
The dark matter distributions of M31 and the Milky Way were assigned an NFW
halo mass profile with a concentration parameter $c=11.9$ \citep{Navarro}.
Our results are not sensitive to reasonable variations in $c$. We also
tested the truncated singular isothermal sphere as an alternative dark
matter mass distribution and found that it produced similar, albeit
slightly greater, tidal stripping and heating of M33.

The stellar disk of M33 at 10 or 5 Gyr ago was simulated with typically 100
collisionless particles uniformly placed between 2.5 and 7.5 kpc of the
center of M33 in the $(x,y)$ plane.  The outer cutoff simulates
approximately the maximum extent of the stellar disk in M33 \citep{Corb}.
The inner cutoff was adopted for computational speed; since tidal heating
and stripping starts from the outside and proceeds inward, the fate of test
stars near the center of M33 is of little interest for our purposes.

Since strong tidal stripping and heating of M33's halo is also a
possibility, the dark matter distribution of M33 was monitored by placing
roughly 100 ``mass-tracing'' particles at 1 kpc spacings out to the
truncation radius of the galaxy.  The enclosed mass profile for M33 was
updated every $10^8$ yr from the locations of the mass-tracing particles,
azimuthally averaged and binned as a function of radius from the center of
M33.  As these mass-tracing particles interacted with M31 and the Milky
Way, some were tidally stripped, steepening the fall-off of mass with
radius in M33.  The halo stripping and heating of M33 did not affect
significantly its orbit (because of its low mass) but had some effect on
the tidal disruption of its stellar disk.  We therefore started the
calculation by integrating the coordinates of the three galaxies backwards
in time to obtain initial conditions either 10 or 5 Gyr ago.  This was
implemented by reversing the sign of today's velocity vector for each
galaxy.  With the entire system specified in the past, the calculation
proceeded forward in time to the present epoch and included the stripping
and heating of M33's dark halo and stellar disk.
%We adopted an NBODY0 softening length ($\sqrt{{\rm eps2}}$) of 0.005~pc and
%an accuracy parameter (``eta'') of $10^{-4}$; test runs with these values
%decreased by an order of magnitude each yielded essentially the same
%results.

\begin{figure*}[t]
\epsscale{0.8} 
\plotone{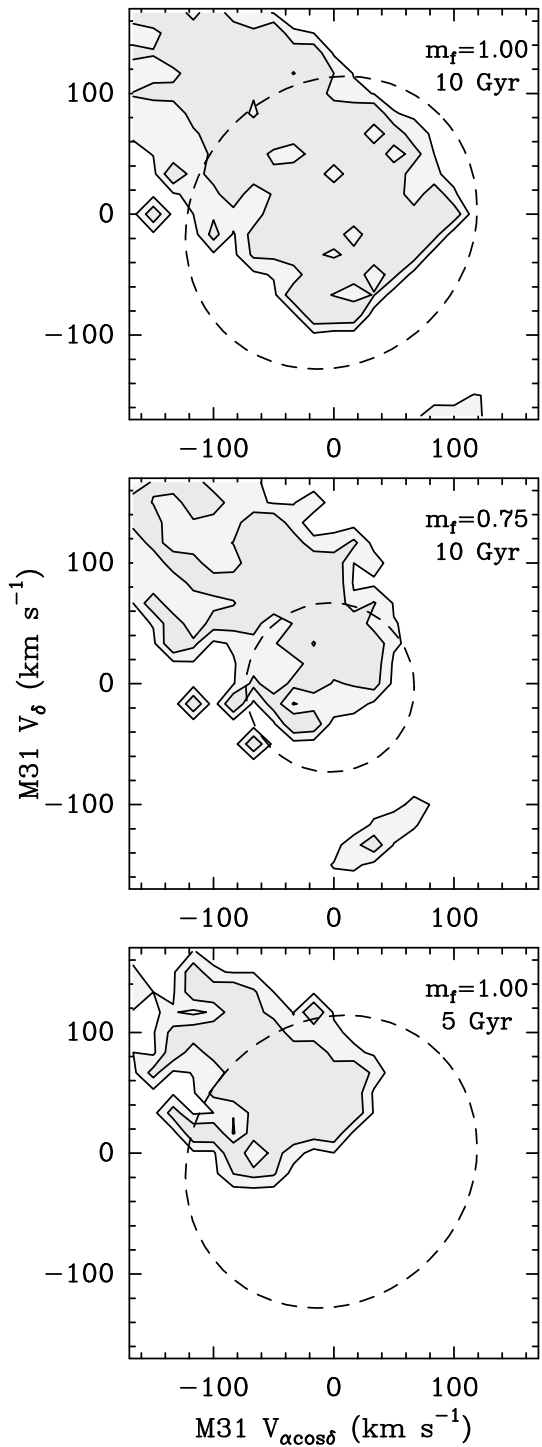}
\caption{\label{fig2} The fraction of tidally stripped stars as a function
of the trial proper motion of M31.  Trial proper motions are in Equatorial
coordinates (Right Ascension, $\alpha$, and Declination, $\delta$).
Contours delineate 20\% (light grey) and 50\% (dark grey) of the total
number of stars stripped in the simulation.  Regions inside the dashed
ellipses correspond to trial proper motions that satisfy the condition that
the separation between M31 and the Milky Way was smaller 10 Gyr ago than it
is today (the ``timing'' argument).  {\it Top:} Simulations with full dark
matter halos ($m_f=1$) and over the past 10 Gyr.  The excluded region of
proper motion is directed North-West.  {\it Middle:} Simulations with
reduced dark matter halos ($m_f=0.75$) and over the past 10 Gyr.  {\it
Bottom:} Simulations with full dark matter halos ($m_f=1$) and over the
past 5 Gyr.  }
\end{figure*}

Once the local group arrived at the present time, we displayed the
locations of the ``disk'' stars in M33, as illustrated in Figure~1.  In
some trials, the disk stars remained in orbit about the center of M33 with
little change in radial distribution or the original orbital plane,
indicating little tidal stripping and heating.  In other trials, many stars
were moved to much greater radii, the mean orbital plane was rotated and
warped, and the distribution was more triaxial than planar, indicating
strong tidal heating after a few passages near M31 or the Milky Way.

\begin{figure*}[t]
\epsscale{1} 
\plotone{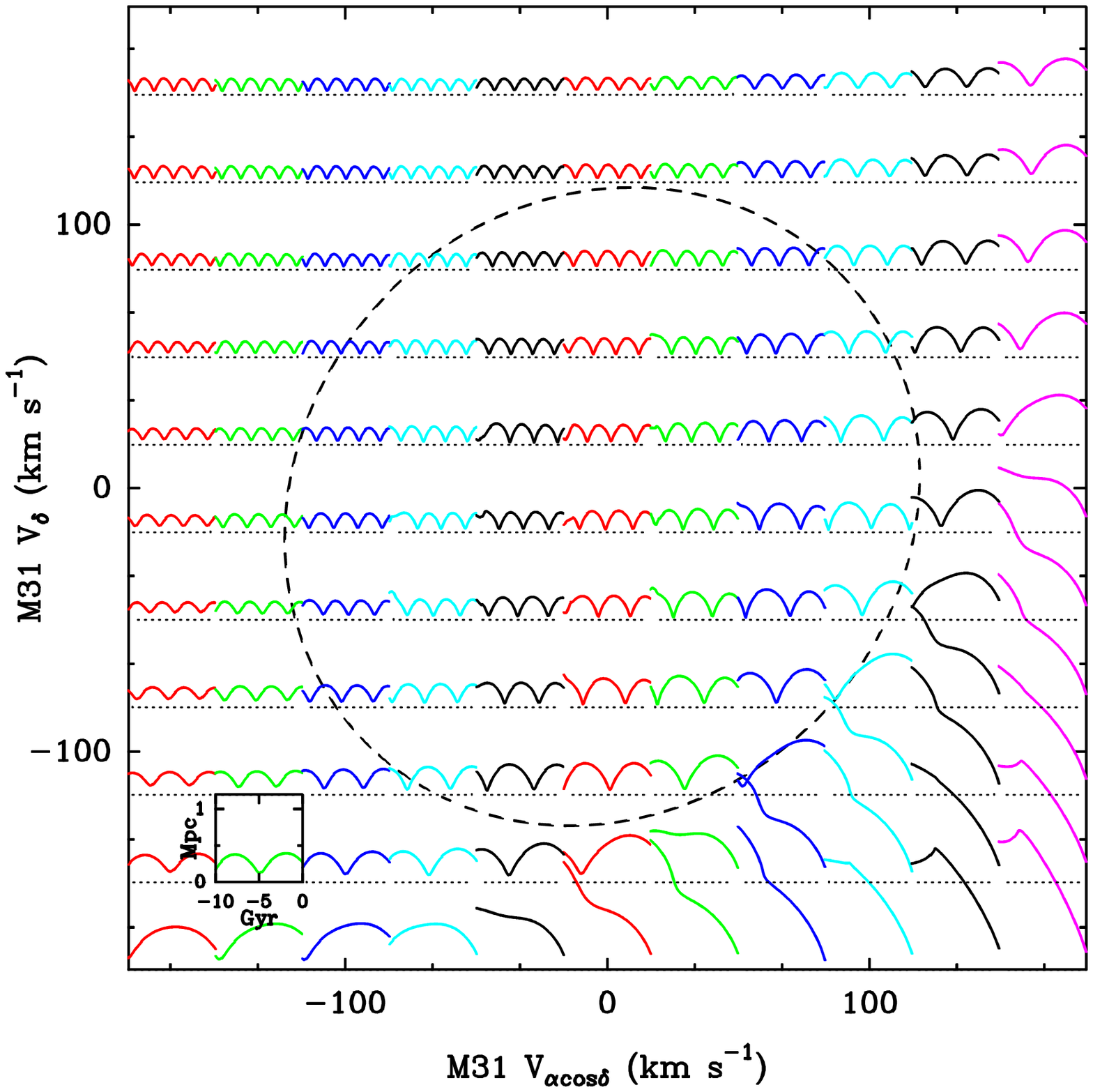}
\caption{
\label{fig3}
The separation between M33 and M31 as a function of lookback time for many
trial proper motions of M31 (indicated the bounding axes).  The scale for
the individual plots is marked on the box near the lower left corner.  The
region inside the dashed ellipse corresponds to trial proper motions that
satisfy the condition that the separation between M31 and the Milky Way was
smaller 10 Gyr ago than it is today.  All examples assume the full dark
matter halos ($m_f=1$).  }
\end{figure*}

In order to evaluate quantitatively the degree of tidal stripping and
heating, we produced a simple estimate of the percentage of stars
``stripped'' from M33.  Stars moved to radii greater than 8 kpc or out of
the initial stellar plane by more than 20\% of their radial distance, were
considered stripped.  We determined the stellar plane in an approximate
manner, by fitting straight lines to the projected distributions in the
$(x,z)$ and $(y,z)$ planes, rotating the system back to the $(x,y)$ plane,
and calculating each star's distance from this plane ($z-$distance).  In
general, three Euler angles are needed to correct the rotation of a system.
However, for small tidal rotations of the stellar plane, our procedure is
adequate since we start with the stars in a plane aligned with the
Cartesian axes.  For large rotations ($\ga 30^\circ$), we optimized this
procedure by considering a $7\times7$ grid of rotation angles spaced by
$7^\circ$ and centered on the initial results; for each test we calculated
the percentage of stripped stars and adopted the rotations that produced
the minimum stripping.

Figure 2 displays the percentage of stars in M33 that were tidally stripped
over a wide range of trial proper motions for M31.  For the galaxy masses
listed above and a lookback time of 10 Gyr, large areas of proper-motion
space for M31 result in significant ($>50$\%) tidal stripping of M33's
stellar disk, including most proper motions smaller than $\sim 75$~\kms.
We repeated our calculations for two lookback times (5 or 10 Gyr) and for
two estimates of total galaxy masses (given by $m_f$ of 1.0 and 0.75).
Reducing either the lookback time or the masses of the galaxies reduces the
stripping, although significant stripping and heating still occurs over
large areas of proper-motion space for M31.  Since the real M33 does not
appear to have undergone significant tidal heating, the darkly shaded areas
in Figure 2 would seem to be ruled out as possible proper motions for M31.

We tested the sensitivity of our conclusions to uncertainties in M33's
proper motion, the relative distances of M31 and M33, and the initial
orientation of the stellar disk in M33.  In one test, shifting the motion
of M33 by 70 \kms\ (randomly in the three Cartesian coordinates) increased
the tidal stripping and heating by about 20\%.  Increasing (decreasing) the
distance of M31 by 34 kpc [$\sim \sqrt{2}$ times the $1\sigma$ precision of
the observations in McConnachie et al. (2004,2005)] also increased
(decreased) the stripping by $\sim 20$\%. The conclusions of the paper
would change significantly only if the distance to M31 is decreased by $\ga
50$ kpc (or through a similar increase in M33's distance, since the tidal
effect depends primarily on the separation between M31 and M33).  Finally,
rotating the original stellar disk of M33 by $90^\circ$ (from the $(x,y)$
to the $(x,z)$ plane) reduced the stripping and heating by about 30\%.
Overall, we conclude that our results are only sensitive at the level of
tens of percent to uncertainties in the initial configuration of our
simulations.
 
One other constraint on the proper motion of M31 results from the
``timing'' argument \citep{Kahn,Peebles2}.  Because the Universe has
expanded considerably over the past $\sim10$~Gyr, galaxies (or
proto-galaxies) should have been closer together in the distant past than
they are today.  Any large proper motion for M31 violates this argument.
Since our calculations produce a time history of the separation of M31 and
the Milky Way, we plotted this separation as a function of time and
determined the approximate range of M31 proper motions for which these
galaxies were closer together 10 Gyr ago than today.  The regions inside
the dashed ellipses in Figure~2 indicate approximately the proper motions
of M31 for which this occurs.  For values of $m_f \leq 0.5$, we have found
that M31 and the Milky Way were never closer together in the past than
today, for any of our trial proper motions of M31.  Since this violates the
timing argument, we do not show values of $m_f\leq 0.5$ in Figure
2.

For the fiducial case of $m_f=1$ and a lookback time of 10 Gyr (top panel),
the allowed proper-motion parameter space delineates an annular arc $\sim
40$~\kms\ wide that starts near $V_{M31}\approx(-90,0)$, goes counter
clockwise through $(0,-100)$ and $(90,0)$, to $(50,80)$~\kms along Right
Ascension and Declination.  Additional small ``islands'' of moderate
stripping and heating exist with smaller likelihood.

Figure 3 shows the separation between M33 and M31 as a function of lookback
time for the various trial proper motions of M31. Since we initiate all
orbits by integrating M33's orbit back in time without stripping and then
integrate these orbits forward in time with stripping, we lose
time-reversal symmetry and so there is some variation in the final
separation between M31 and M33 among the different cases.  Throughout the
entire range allowed by the timing argument (delineated by the dashed
line), M33 orbits around M31 several times as expected from a
gravitationally-bound satellite.  The repeated encounters, while not
disruptive in the allowed range of proper motions, might have still
stripped material from the outskirts of M33 and could potentially account
for the streams of \ion{H}{1} gas \citep{Braun1,Braun2} and stars
\citep{Guha} that are observed in between the two galaxies today.

\section{Conclusions}

Figure 2 presents our limits on the proper motion of M31 based on the
constraint that the thin stellar disk of M33 (simulated as an ensemble of
collisionless test particles on circular orbits) was not tidally disrupted
over the past $10^{10}$ years through its orbit within the local group.
This constraint and the condition that M31 and the Milky Way galaxies were
closer in the past, favors a proper motion amplitude of $100\pm20 ~{\rm
km~s^{-1}}$ for M31 with the quadrant of a negative velocity component
along Right Ascension and a positive component along Declination strongly
ruled-out. The orbital angular momentum associated with the allowed range
of proper motions implies that the dark halos of Andromeda and the Milky
Way will pass through each other within the next 5--10 billion years.  The
main uncertainty in our results involves the dark halo masses of the local
group members (compare the top two panels in Fig. 2).
 
Our inference relies on the validity of various simplifying assumptions.
For example, we simulated the dynamics of M33 without taking account of the
full dark-matter distribution interior to the local group (including the
mass in between the Milky Way and Andromeda). In addition, we have assumed
that the tidal influence of external objects on the local group can be
neglected, and we have ignored the addition of mass to the local group
through accretion and infall of external matter.  Obviously, the Milky Way
and the Andromeda galaxies grew hierarchically with cosmic time but the
ages and low velocity dispersion of stars within their disks argue against
major merging encounters over the past 10 billion years
\citep{Wyse1,Wyse2}.  Our quantitative predictions provide an added
incentive for future astrometric observatories such as
SIM\footnote{http://planetquest.jpl.nasa.gov/SIM/sim\_index.html}, GAIA
\footnote{http://sci.esa.int/science-e/www/area/index.cfm?fareaid$=$26}, or
the SKA\footnote{http://www.skatelescope.org/}, that will be able to test
them \citep{Shaya}.

\acknowledgements

We thank Andy Gould and Robert Braun for useful comments.  This work was
supported in part by NASA grant NAG 5-13292, and by NSF grants AST-0071019,
AST-0204514 (for A.L.).

\end{document}